\PassOptionsToPackage{dvipsnames}{xcolor}
\documentclass{aa}  

\usepackage[normalem]{ulem}
\usepackage{listings}
\usepackage{graphicx}
\usepackage{amsmath}
\usepackage{natbib}
\usepackage[dvipsnames]{xcolor}
\usepackage{array,booktabs}
\usepackage{multirow}
\usepackage{txfonts}
\newcommand{\kms}{\,\hbox{\hbox{km}\,\hbox{s}$^{-1}$}}

\begin{document}

   \title{Andromeda's tenuous veil: A likely Milky Way nebula projected toward M31}

\author{A. Lumbreras-Calle
        \inst{\ref{CEFCA}}\fnmsep\thanks{\email{alumbrerascalle@gmail.com}}
       \and J.~A.~Fern\'andez-Ontiveros\inst{\ref{CEFCA},\ref{UA}}
 \and R.~Infante-Sainz\inst{\ref{CEFCA}}
\and M.~Akhlaghi\inst{\ref{CEFCA},\ref{UA}}
        \and B.~Montoro-Molina\inst{\ref{IAA},\ref{Chalmers}} 
         \and B.~Pérez-Díaz\inst{\ref{IAA}}  
        \and A. del Pino\inst{\ref{CEFCA},\ref{UA},\ref{IAA}}
 \and H.~Vives-Arias\inst{\ref{CEFCA}} 
 \and A.~Hern\'an-Caballero\inst{\ref{CEFCA},\ref{UA}}
 \and C.~L\'opez-Sanjuan\inst{\ref{CEFCA},\ref{UA}}
\and M.~A.~Guerrero\inst{\ref{IAA}} 
\and S. Eskandarlou\inst{\ref{CEFCA}} 
       \and A.~Ederoclite\inst{\ref{CEFCA},\ref{UA}}
}
\institute{Centro de Estudios de F\'isica del Cosmos de Arag\'on (CEFCA), Plaza San Juan 1, 44001 Teruel, Spain\label{CEFCA}
        \and
        Unidad Asociada CEFCA--IAA, CEFCA, Unidad Asociada al CSIC por el IAA, Plaza San Juan 1, 44001 Teruel, Spain\label{UA}
        \and
        Instituto de Astrof\'isica de Andaluc\'ia (IAA-CSIC), P.O.~Box 3004, 18080 Granada, Spain\label{IAA}
        \and
        Department of Space, Earth and Environment, Chalmers University of Technology, Onsala Space Observatory, 43992 Onsala, Sweden\label{Chalmers}
        }
  \date{}

  \abstract
{A large, faint nebula was unexpectedly discovered near M31 using narrowband [O~III] images. Its apparent size and the lack of a clear counterpart at other wavelengths make it unique and challenging to explain.}
{We aim to determine whether the nebula is extragalactic and vast or associated with the Milky Way filamentary structure. This will enable us to constrain its physical properties and assess its nature.}
{We obtained deep narrowband [O~II]3727 and H$\alpha$+[NII] observations with the JAST80 telescope at the Observatorio Astrofísico de Javalambre, as well as high spectral resolution spectroscopy (R$\sim$5000) at four locations within the region of interest using the MEGARA integral field unit at the Gran Telescopio Canarias.}
{We found extended [O~II] emission along two near-parallel strands to the [O~III], offset by six arcmin. The nebular spectra reveal up to six emission lines from [O~III]4959,5007, H$\beta$, [N~II]6583, and [S~II]6716,6731. Their receding velocities are above $-$40 km s$^{-1}$, far from the systemic velocity of M31 ($-$300 km s$^{-1}$). The fluxes and velocities are remarkably consistent for the same lines across different regions of the nebula.} 
{The nebular properties suggest a location within the Milky Way rather than being physically associated with M31. The most likely scenario suggests a resolved ionization structure in a Galactic nebula with a separation between [O~II] and [O~III] on the order of a few parsecs. The observed receding velocities would be unprecedented for an object physically linked to M31 but are common for nearby gas filaments. Their consistency across the nebula would also be highly unusual if it were larger than a kiloparsec. The analysis of the emission line ratios, line widths, and morphology suggests the possibility of it being an interstellar gas filament with an additional source of ionization to explain the [O~III] emission. However, the complex properties of this object call for further observations to confirm its nature.}


   \maketitle

\section{Introduction}

Since the middle of the 20th century, in addition to broadband imaging, the sky has been observed with surveys focused on studying nebular line emission. Early examples were blind surveys performed with objective prisms, unveiling large numbers of Galactic objects such as planetary nebulae and supernova remnants \citep{Minkowski46}, a technique also applied to the discovery of extragalactic objects such as emission line galaxies \citep{Markaryan,MacAlpine}. Narrowband photometry was also developed early, mostly focused on H$\alpha$ observations of Galactic objects \citep{Shajn,Gum} but also nearby galaxies \citep{Shajn,Courtes}. Many more surveys, focusing on H$\alpha$ within our own Galaxy, have continued this work, such as the Virginia Tech Spectral-line Survey (VTSS; \citealt{VTSS}) and IPHAS \citep{IPHAS}. One early example of a Galactic survey targeting other optical emission lines was \cite{Parker}, leading to the discovery of multiple planetary nebulae. The Wisconsin H$\alpha$ Mapper survey (WHAM; \citealt{WHAM}) has primarily observed spectroscopically in the H$\alpha$ wavelength range, with some fields covered in other emission lines, providing high spectral resolution over wide areas of the sky.   

Beyond our galaxy, the systematic surveys performed over the past two decades using narrowband filters have only targeted small areas of the sky, mainly focusing on compact objects at intermediate and high redshifts (i.e., \citealt{Cardamone,Moles08,pg13,Taniguchi}). Using large-scale, broadband surveys, only a few extreme examples of extended, ionized structures beyond the local group have been discovered, the most notable being extended emission line regions (EELRs; \citealt{Lintott09,Keel12}). With a stronger discovery potential for these kinds of structures, several recent narrowband projects are targeting large sky areas, such as the Merian survey \citep[][submitted to ApJ]{Danieli}, the Javalambre Photometric Local Universe Survey (J-PLUS; \citealt{Cenarro19C}), the Southern Photometric Local Universe Survey (S-PLUS; \citealt{splus}), and the Javalambre Physics of the Accelerated Universe Survey (J-PAS; \citealt{Benitez14,Bonoli}). They have already yielded results, for example, on extended Lyman-$\alpha$ emission around quasars \citep{Rahna}, jellyfish galaxies \citep{Gondhalekar}, and nearby extreme emission line galaxies \citep{Lumbreras}.

Despite the long history of observations of emission line objects, there remains a large potential for new discoveries. The low surface brightness (LSB) regime has been poorly explored due to observational constraints, especially targeting emission lines. In the more distant Universe, integral field unit (IFU) spectroscopy has been used in the past decade to measure faint emission line structures. Using the Multi-Unit Spectroscopic Explorer (MUSE) IFU, with data from the MUSE Ultra Deep Field \citep{muse_hdf}, extended Lyman-$\alpha$ emission was identified around most distant galaxies \citep{muse_lyman}, and it was also found to trace the cosmic web structure \citep{muse_cosmicweb}. Using stacking techniques, deep MUSE data were used as well to detect H$\alpha$-emitting sources near galaxies (likely intermediate-mass black holes) or to measure the extended line emission around galaxies \citep{Dutta}. 

In the nearby Universe, however, there is still a lack of deep, wide observations of emission lines over wide fields. Amateur astronomers, targeting relatively wide areas of the sky for hundreds of hours, are able to obtain deeper observations than what is available in the scientific literature for specific fields. One example of these projects is the one leading up to the discovery of a large nebula near M31. It was detected by amateur astronomers observing with a narrowband [O~III]5007 filter \citep{Drechsler23}, and they named it Strottner-Drechsler-Sainty Object 1 (SDSO-1), hereafter SDSO. The nebula is very faint, with a [O~III] surface brightness of $\sim$ 10$^{-18}$ erg s$^{-1}$ cm$^{-2}$ arcsec$^{-2}$, yet it is very large ($\sim$ 1.5 deg) and presents a filamentary structure. One of its most striking features was the apparent lack of any counterpart, neither in H$\alpha$ emission nor in X-ray, UV, optical, or infrared observations. In \cite{Fesen23}, the photometric data are described in more detail, and additional spectroscopic follow-up is presented, using the Ohio State Multi-Object Spectrograph (OSMOS) at the MDM 2.4m Hiltner telescope. It showed a receding velocity of $\sim-$10 km s$^{-1}$. Despite this, in an extensive and detailed analysis, they conclude that the nebula is unlikely to be located within the Milky Way by ruling out its classification as several classes of Galactic objects. They favor the view that the nebula is instead in the M31 halo, ionized by shocks. \cite{Amram23} also performed spectroscopic observations with MISTRAL at the 193 cm telescope of the Haute-Provence Observatory, with discrepant results (a receding velocity of $\sim~-$100 km s$^{-1}$). This paper was prepared before the submission of a recent Arxiv preprint \citep[][submitted to ApJ]{Ogle} featuring deeper narrowband observations and a new Galactic model for SDSO. The authors propose that the object is a ghost planetary nebula produced by the symbiotic star EG\,Andromedae, which lies in the direction of M31 at a distance of 60 pc from the Sun.

In this work, we present additional photometric and spectroscopic observations of the nebula. We discuss possible scenarios for its location and physical nature, comparing SDSO with different classes of objects and taking into account the available data. This paper is structured as follows: In Sect. \ref{sec:databas} we describe our observations. In Sect. \ref{sec:results}, we present the direct results of the observations and the measurements we performed on them. In Sect. \ref{sec:discussion}, we discuss the implications of the results, comparing them with the literature to infer the most likely location and physical nature of the nebula. In Sect. \ref{sec:conclusions} we summarize our conclusions.

\section{Observational data}
We secured follow-up observations on the M31 [O~III] nebula to constrain its possible physical properties and determine its Galactic or extragalactic nature. More specifically, we obtained narrowband photometry targeting both the [O~II] and H$\alpha$ emission lines, and integral field spectroscopy covering the ranges from H$\beta$ to [O~III]5007 and from H$\alpha$ to [S~II]6731. In this section we present each of these datasets within their respective subsections.

\label{sec:databas}

\subsection{OAJ photometric observations}

\label{sec:oaj_photom_data}
\begin{figure*}
   \centering
   \includegraphics[width=0.98\textwidth,keepaspectratio]{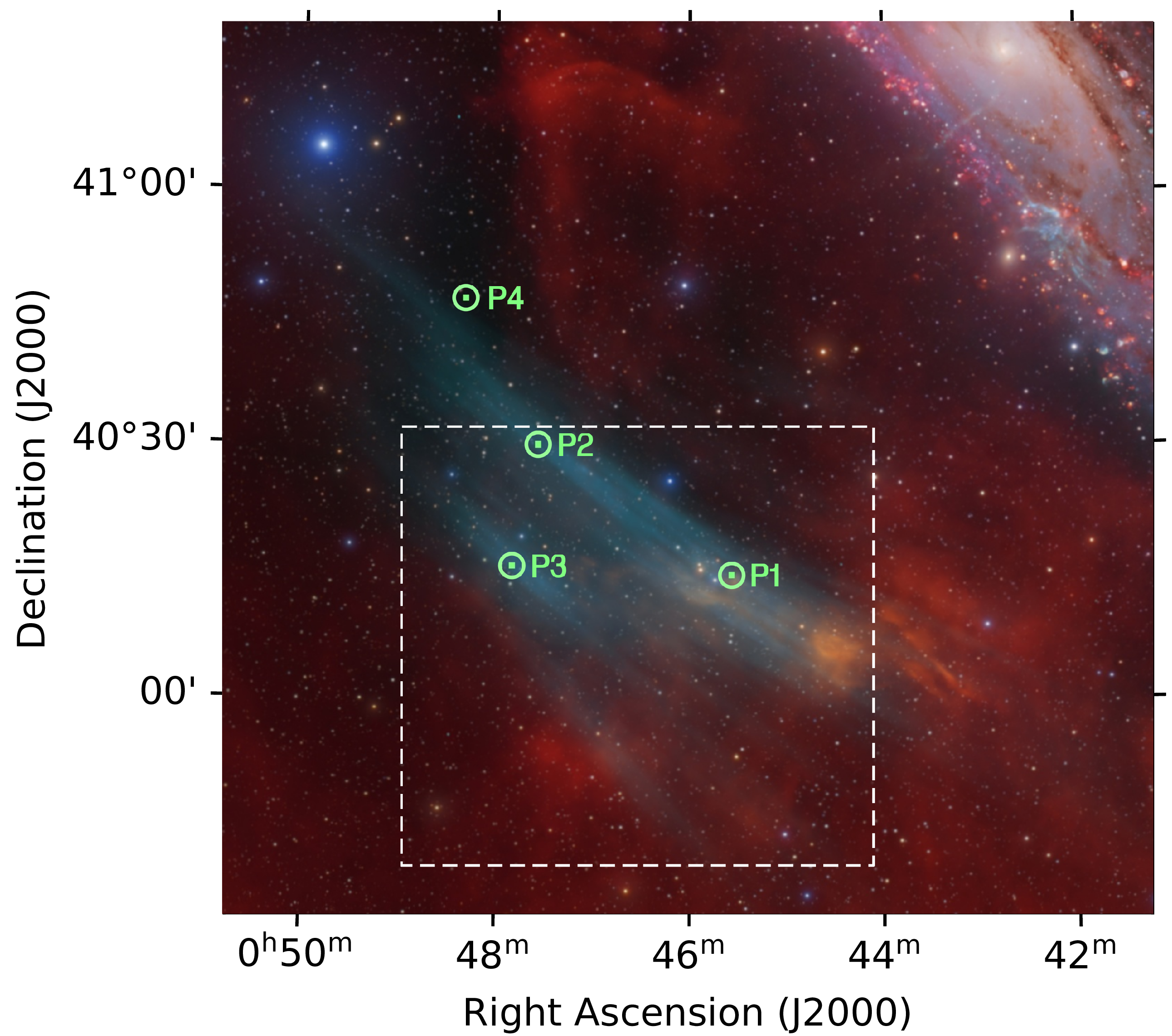}
      \caption{Image of the [O~III] emitting nebula (known as SDSO) and its surrounding area, including the Andromeda galaxy (top right). The [O~III] flux is shown in cyan, while H$\alpha$+[N~II] is shown in red. The green squares (surrounded by green circles) mark the location of the MEGARA pointings in and around the nebula presented in this work. The dashed square shows the area covered by our full-depth observations with the JAST80 telescope targeting [O~II] and H$\alpha$ with narrowbands $J0378$ and $J0660$, respectively. The RGB image\protect\footnotemark of the nebula and M31 was presented in \cite{Drechsler23}, and was obtained by Marcel Dreschler, Xavier Strottner, and Yann Sainty.}
         \label{fig:pointings}
   \end{figure*}

In order to explore the morphology of the nebula in other optical emission lines, we obtained photometric observations with the Javalambre Auxiliary Survey Telescope (JAST80), an 80 cm telescope located at the Observatorio Astrofísico de Javalambre (OAJ), thanks to its Director's Discretionary Time (DDT) program, through proposal 2300222 (P.I. Lumbreras-Calle). The camera (T80Cam) is optimal for wide-field imaging, thanks to its 9.2k $\times$ 9.2k CCD, with a pixel scale of 0.55 arcsec pix$^{-1}$, resulting in a Field of View (FoV) of 1.4 deg $\times$ 1.4 deg. Its primary task is to perform J-PLUS \citep{Cenarro19C}, a 8500 deg$^2$ photometric multiband survey using a set of 12 filters (five broad and seven narrow or medium).

Considering the available filters on the T80cam in JAST80, we proposed observations using the $J0378$ filter ($\lambda$$_{eff}$=3793.4~\AA, $W_{eff}$=136.3 \AA~) which covers the [O~II]3727 emission line. This line is key in the cooling of a nebula and in constraining its physical properties. We also observed the nebula with the $J0660$ filter ($\lambda$$_{eff}$=6606.9~\AA, $W_{eff}$=146.6~\AA), covering both the H$\alpha$ and [N~II]6548,6583 emission lines. This allows us to compare our JAST80 images with the existing observations in \cite{Drechsler23,Fesen23} and our spectra.

The detailed account of the method we followed to observe and process the images is presented in Appendix \ref{app:data}. In short, we observed multiple times with a large dithering that resulted in combined images with total exposures of 1.5 and 3.67 h. The data reduction involved the usual J-PLUS pipeline, as well as multiple tools, such as the \texttt{NoiseChisel} program \citep{akhlaghi2015,akhlaghi2019}, and several techniques optimized to recover low surface brightness structures (e.g., removing contamination from stars following \citealt{infantesainz2020}.)

\footnotetext{\url{https://www.astrobin.com/1d8ivk/0/}}

\begin{figure}
\includegraphics[width=0.45\textwidth,keepaspectratio]{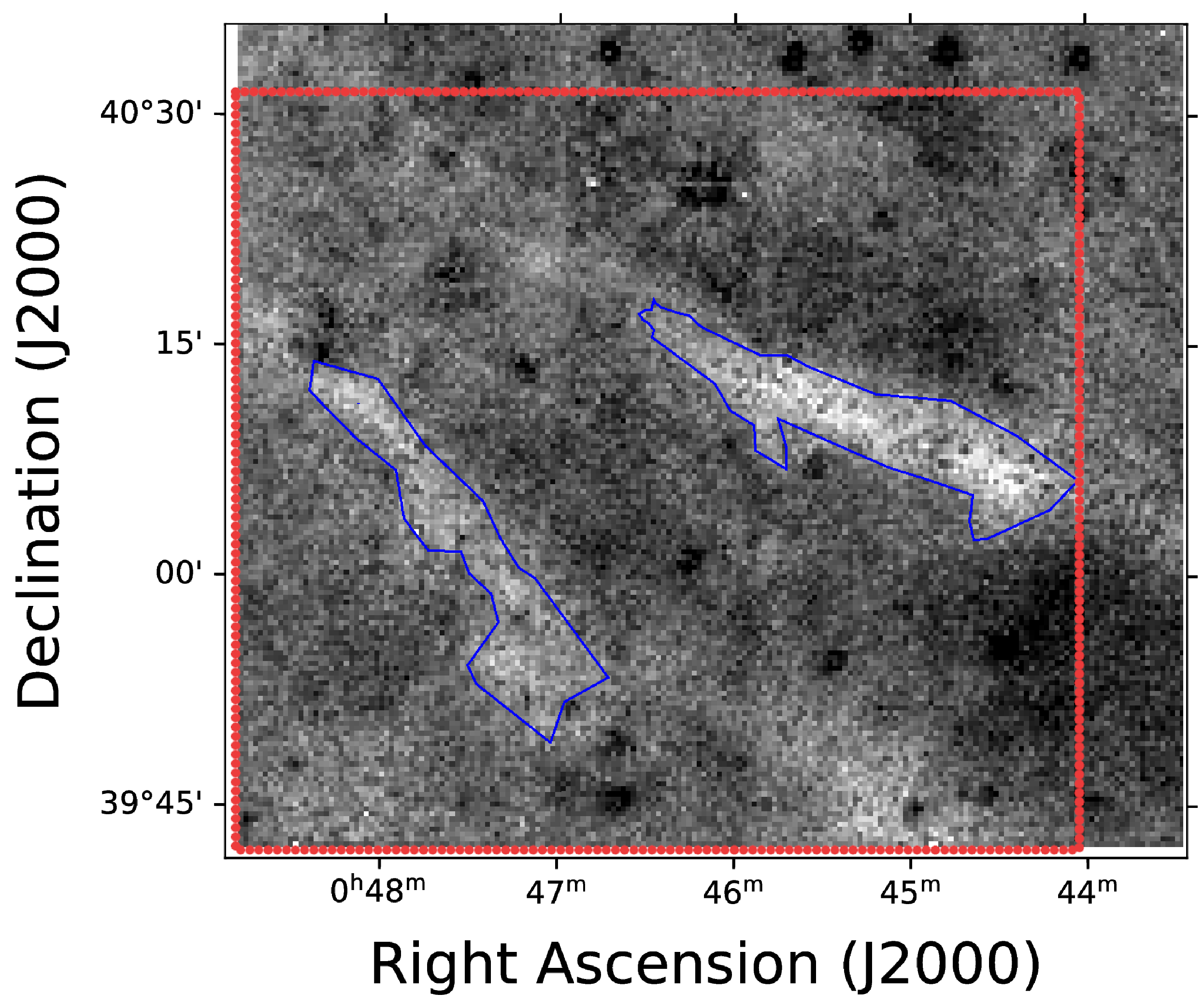}
\includegraphics[width=0.45\textwidth,keepaspectratio]{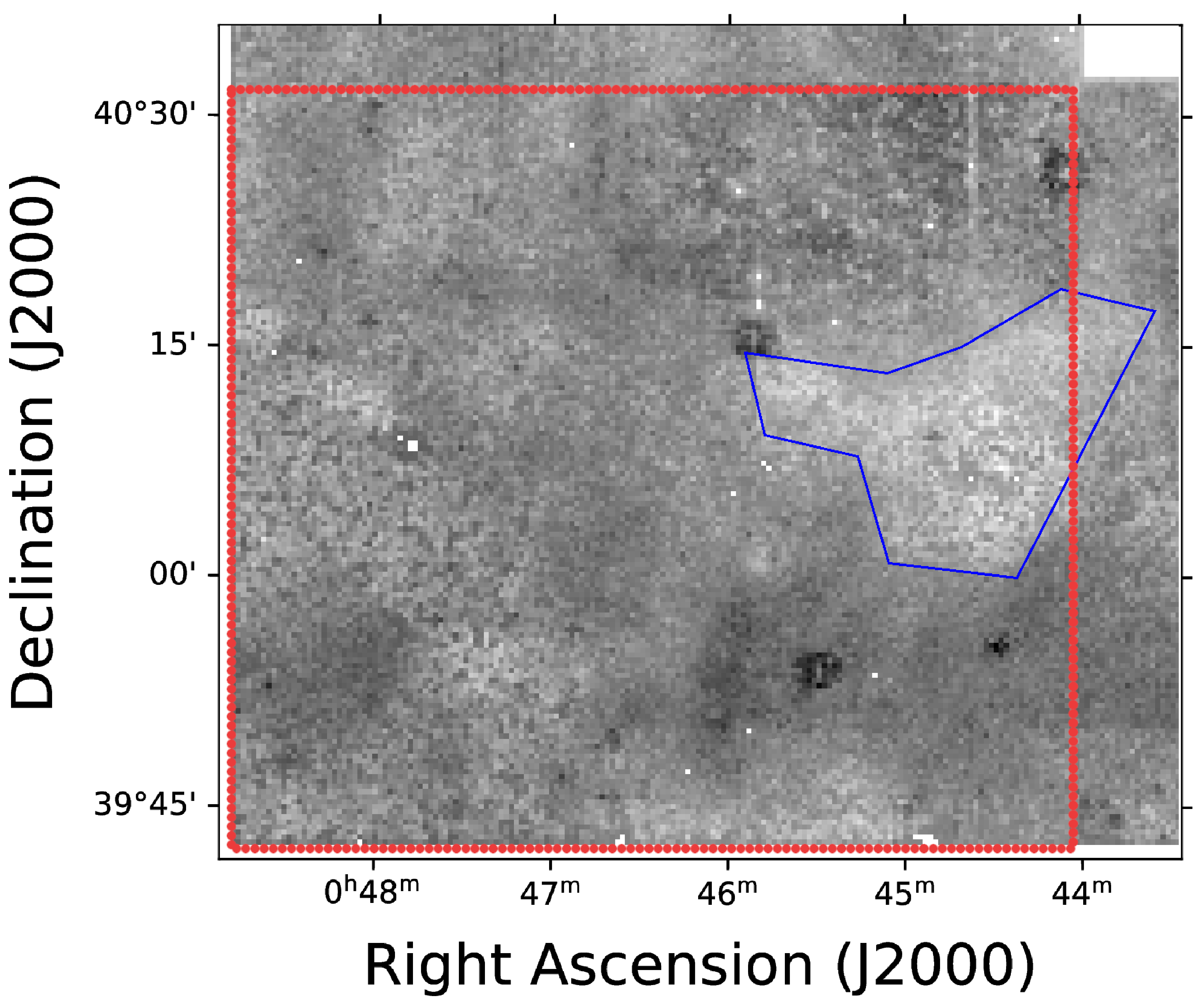}
      \caption{Narrowband images obtained using JAST80 on the [O~III] nebula. \textit{Top.} $J0378$ image, tracing the [O~II]3727 emission line. \textit{Bottom.} $J0660$ image, tracing H$\alpha$ and [N~II] emission. In both images, in blue, we show the contours of areas where we detect emission in each image. In dotted, red lines, we encircle the area where the full depth in the images is achieved.}
         \label{fig:oaj_photom_data}
   \end{figure}

The final results after the data reduction process are shown in Fig. \ref{fig:oaj_photom_data}, after rebinning the images to increase the pixel scale up to 20\ arcsec pix$^{-1}$. We have highlighted only the region where we reach the full depth, a rectangle of 56 arcmin by 50 arcmin of size, giving a total area of 2750 arcmin$^2$ (0.75 deg$^2$).

\subsection{MEGARA spectroscopic observations}
\label{sec:megara_sec2}

\begin{figure*}
   \centering
   \includegraphics[width=0.98\textwidth,keepaspectratio]{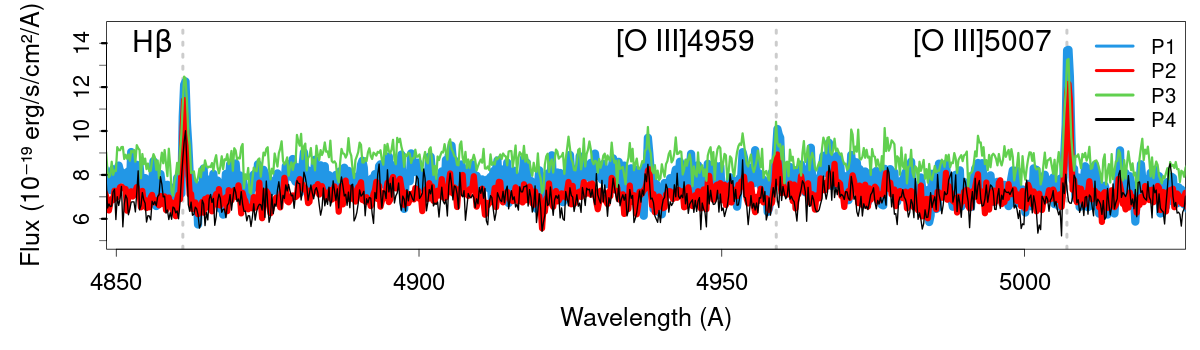}
   \includegraphics[width=0.98\textwidth,keepaspectratio]{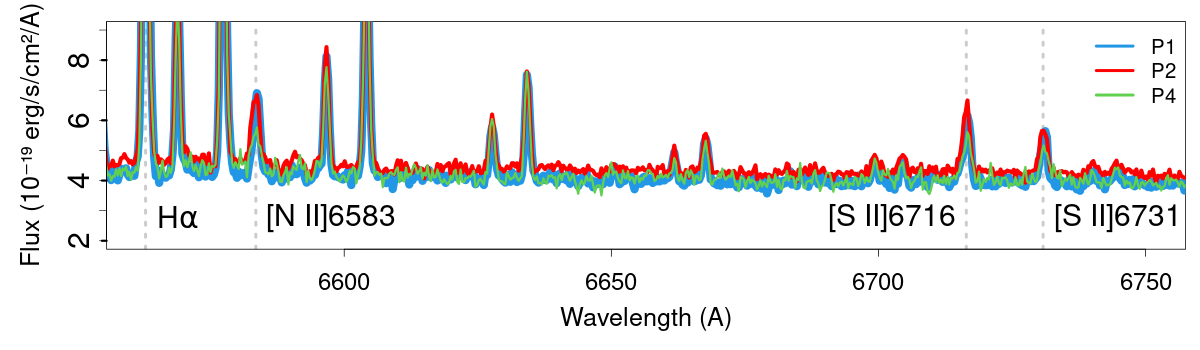}
      \caption{MEGARA spectra of the pointings in the first night (four in the blue range, three in the red), in the observed wavelength frame. \textit{Top.} The LR-B spectra, centered in the wavelength range from H$\beta$ to [O~III] 5007. \textit{Bottom.} The LR-R spectra, from [N~II] 6583 to [S~II]6731. The emission lines of interest are marked in both plots. The lines that are not marked are sky emission lines, with H$\alpha$ being contaminated by one.}
         \label{fig:examspec}
   \end{figure*}
To accurately measure the receding velocity of the nebula and assess its physical state, we obtained spectroscopic observations using the IFU mode of the Multi-Espectrógrafo en GTC de Alta Resolución para Astronomía (MEGARA) instrument mounted at the 10.4-m Gran Telescopio Canarias (GTC) in the Observatorio del Roque de los Muchachos (ORM). We obtained observing time through the DDT program GTC2023-206 (P.I. Lumbreras-Calle). We targeted three different regions in the nebula (P1, P2, and P3) and one outside of its apparent footprint (P4; see Fig. \ref{fig:pointings}). The pointings P1 and P2 were selected along the major axis of the object, focusing on regions with significant [O~III] emission, and at least one (P1) where the H$\alpha$ emission was noticeable as well. P3 was selected in a separate “strand” of the nebula, along what could be considered its minor axis. The precise central coordinates of the pointings are listed in Table \ref{tab:coords} and were chosen to avoid any other visible source within the MEGARA IFU footprint  (approximately 12.5 $\times$ 11.3 arcsec) to minimize contamination.

\begin{table}[t] 
\caption{Central position of the MEGARA pointings. }
\label{tab:coords}
\centering 
        \begin{tabular}{c c c } 
        \hline\hline 
       Pointing  & RA     &  DEC       \\ 
                & [deg]   &  [deg]  \\ 
        \hline
        P1 & 11.3792     &   40.2435       \\ 
         P2 & 11.8913 & 40.4936          \\  
         P3 &  11.9480 & 40.25428               \\  
        P4 &  12.0784 & 40.7820              \\ 
        \hline 
\end{tabular}
 
\end{table}

To increase the signal-to-noise ratio (S/N) in the spectra, and to potentially detect multiple emission lines to determine physical properties, we used the lowest resolution dispersion elements available in MEGARA, since they would still be enough to perform accurate velocity measurements (R $\geq$ 5000). The Volume Phase Holographic (VPH) dispersion elements we used to cover the wavelength range of [O~III] was VPH480-LR (LR-B), ranging from 4332.05~\AA~ to 5199.96~\AA~, with a pixel scale of 0.207~\AA~ pix$^{-1}$. In addition, we observed pointings P1, P2, and P4 with VPH675-LR (LR-R), ranging from 6096.54~\AA~ to 7303.21~\AA~ and a pixel scale of 0.287~\AA~ pix$^{-1}$, to cover the wavelength range of H$\alpha$, [N~II] and [S~II]. Distinguishing between different ionizing sources requires the simultaneous analysis of both highly-ionized ions (e.g., [O~III]) and low-ionized species (e.g., [N~II], [S~II]). Additionally, the information from the [S~II] doublet is needed to constrain the electron density of the nebula.

The observations were carried out over two nights, 2023/01/21 and 2023/01/23. In each night, the same observation sequence was performed. For pointings P1 and P2, three 600 s exposures in each VPH were taken. In P3, only observations in LR-B were taken, and in P4, only 1 exposure with each VPH was taken. Since the object is much more extended than the MEGARA IFU, we placed no constraints on the seeing, and the observations were taken under filler conditions.

Using the observations of a standard star each night, we performed flux calibration with the MEGARA pipeline. The final product of the data reduction was therefore a flux-calibrated 2D spectrum with one row per fiber. The main science-ready output we extracted from it was a 1D spectrum obtained by taking the median of all rows. The only correction at this stage was to remove the sky fibers from the median to avoid mixing different regions of the nebula. The final reduced spectra for all pointings in the first night and both VPHs are shown in Fig. \ref{fig:examspec}, focusing on the wavelength ranges of interest in this work.

\section{Results}
\label{sec:results}
In this section we describe the outcome of our observations. We first describe the results from the JAST80 imaging in two narrowbands. Then we present the results of the spectroscopic observations targeting the nebula, describing the velocity, width, and flux of the different emission lines.

\subsection{Photometry}
\label{sec:result_photom}

After performing the data reduction of the OAJ data (presented in \ref{sec:oaj_photom_data}), we obtained the images in Fig. \ref{fig:oaj_photom_data}. The top panel shows the image taken with the $J0378$ filter, tracing the emission from the [O~II] 3727 doublet, while the bottom panel image corresponds to the $J0660$ filter and shows H$\alpha$ and [N~II]6548,6586 emission. In both of the images we successfully detect low surface brightness structures in the area where the [O~III] nebula was detected. 

In the [O~II] image (see the top panel in Fig. \ref{fig:oaj_photom_data}), we can clearly identify two separate elongated structures. The brighter one, on the west, is approximately 4 arcmin wide and 30 arcmin long, whereas the fainter one, on the southeast, is slightly smaller (27 arcmin) and thinner (3.7 arcmin). For the brighter one, the median surface brightness is 27.8 mag arcsec$^{-2}$ (7.8 $\times$ 10$^{-18}$ erg s$^{-1}$ cm$^{-2}$ arcsec$^{-2}$, assuming all flux comes from the emission line), with a total integrated magnitude of 13.6 mag. The fainter one presents 28.0 mag arcsec$^{-2}$ (6.5 $\times$ 10$^{-18}$ erg s$^{-1}$ cm$^{-2}$ arcsec$^{-2}$) and 14 mag. For simplicity, we have performed our measurements on manually defined apertures, but moderate changes would not affect the results significantly (for example, a 60\% increase in the size of the lower region would result in just a 5\% decrease in surface brightness). Since these structures are very faint, we pushed the images to their surface brightness limits, and we needed special care to estimate the uncertainties (see the technical discussion in Appendix \ref{app:photom_anal}). This analysis confirmed the significance of the structures detected in [O~II], with a minimum S/N of 5.1.

In the H$\alpha$ image we detect a roughly triangular region of $\sim$ 24-27 arcmin in size. The median surface brightness in that structure is 27.05 mag arcsec$^{-2}$ (5.5 $\times$ 10$^{-18}$ erg s$^{-1}$ cm$^{-2}$ arcsec$^{-2}$). With the same method explained in Appendix \ref{app:photom_anal}, the S/N ranges from 5.3 to 6.7, confirming the reality of this structure.

\subsection{Spectroscopy}

To extract physical information from the spectra, we ran a custom-made code to measure the key properties of the emission lines. After visual inspection, we decided to focus only on the clearly detected lines: H$\beta$, [\ion{O}{iii}]$\lambda4959$, [\ion{O}{iii}]$\lambda5007$ in LR-B, and [\ion{N}{ii}]$\lambda\lambda6548, 6583$, H$\alpha$, and [\ion{S}{ii}]$\lambda\lambda6716, 6731$. As a reference for calibration, we also measured the four strongest sky emission lines found closer in wavelength to the nebular emission lines. The code estimated the continuum level around the emission line by masking the feature and fitting a first-degree polynomial. A Gaussian function was fitted to the line profiles in the continuum-subtracted spectrum to derive the flux, central wavelength, velocity with respect to the rest-frame wavelength, and line width. Line fluxes and best-fit parameters are shown in Table \ref{tab:veloc_flux}. The upper limits correspond to $3\sigma$, where $\sigma$ is the associated uncertainty of the fit. Further corrections applied to the output values of the code are explained in the following subsections and in Appendix \ref{app:spec}. After this procedure, we successfully detect the [\ion{O}{iii}]$\lambda5007$ and H$\beta$ emission lines in all our MEGARA pointings in the nebula (P1, P2, and P3), confirming the initial result in \cite{Drechsler23} (Fig.~\ref{fig:examspec}). 

\subsubsection{Emission line fluxes}
\label{sec:results_emlineflux}
\begin{figure}
\centering
\includegraphics[width=0.5\textwidth,keepaspectratio]{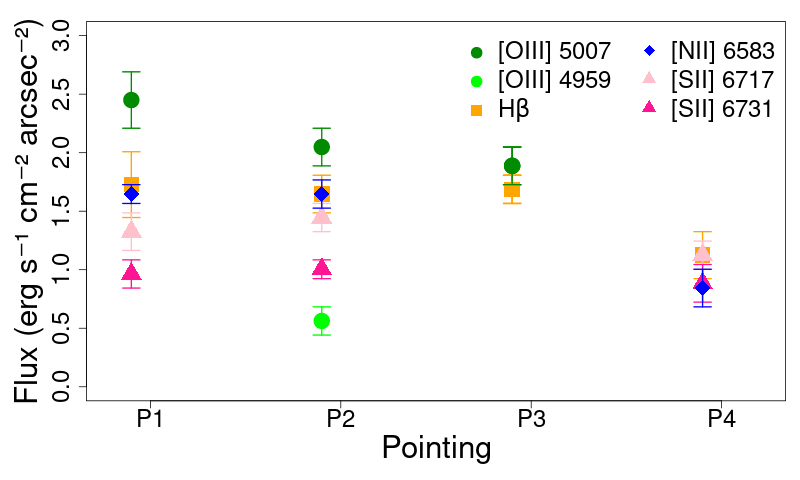}
      \caption{Line fluxes measured with MEGARA at GTC in the different pointings in the region of the nebula on the first night of observations. They correspond to the integrated flux of the best-fitting Gaussian function to the line profile. The lines that are missing have not been detected by our code in the spectra, except for the red lines ($\lambda$>6000 \AA) in P3, where those observations were not made.}
         \label{fig:fluxes}
   \end{figure}

The flux-calibrated 1D spectra obtained for each pointing and night described in Sect. \ref{sec:megara_sec2} were the median of all fibers. To transform it into surface brightness, we divided the flux by the area of one fiber (0.2497 arcsec$^2$, since it is a hexagon with a maximum diameter of 0.62 arcsec). These fluxes are presented in Fig. \ref{fig:fluxes} and Table \ref{tab:veloc_flux}. We excluded from our analysis (and thus from Fig. \ref{fig:fluxes}) the line fluxes measured in the second night, since no flatfield observations were performed. 

We consider that the absolute flux values may hold uncertainties unaccounted for in the stated error intervals. Performing such a calibration over large wavelength ranges is sometimes challenging in spectra, but in this instance it was more complex given the unusual instrument mode (integrating all flux from the 587 fibers), and the very faint and extended nature of the object. Nevertheless, the calibration shows consistent results between pointings, nights, and across different observations.

However, we cannot separate the H$\alpha$ flux in our pointings from the very strong sky emission line at 6562.76~\AA. The flux of this line varies from one pointing and night to the other, in contrast with other sky emission lines, and that variation is likely attributable to H$\alpha$ from the nebula. Nevertheless, the variations are not consistent with the distribution of H$\alpha$ flux in the images, or from the first night to the second. Any conclusion drawn from what is already a heavily contaminated line would be very uncertain.

Finally, we can also compare some of our flux measurements with previous observations of this nebula. The [O~III]5007 emission line flux is consistent with what was estimated by \cite{Drechsler23} in the photometry and by \cite{Fesen23} in both photometry and spectroscopy. More specifically, in our aperture P1 (located only 90 arcsec to the west of Slit 2 in \citealt{Fesen23}) we measure a line surface brightness of (2.45 $\pm$ 0.24) $\times$ 10$^{-18}$ erg s$^{-1}$ cm$^{-2}$ arcsec$^{-2}$, compatible with their measurement of (4 $\pm$ 2) $\times$ 10$^{-18}$ erg s$^{-1}$ cm$^{-2}$ arcsec$^{-2}$.

We find a stronger H$\beta$ emission than what \cite{Fesen23} find in their Slit 2, where it seems roughly [O~III]5007/H$\beta$ $\sim$ 2.5, and our closer pointing P1 shows [O~III]5007/H$\beta$=1.42. This difference may be due to the physical locations explored being slightly different, considering also that relatively large areas are being integrated together, and that the S/N is relatively low.

Our measurements for the [O~III]5007/H$\beta$ ratio in the different locations (or even the spectrum in Slit 2 in \citealt{Fesen23}) are in clear contrast with the photometry-based estimations of [O~III]/H$\alpha>$5 \citep{Drechsler23,Fesen23}. Since we lack spectroscopic measurements of H$\alpha$, we adopt an H$\alpha$/H$\beta$ ratio of 2.86, corresponding to Case B recombination with $T_{e}$= 10\,000 K and $n_e = 300\, \rm{cm^{-3}}$ \citep{Osterbrock06,Luridiana15}. Considering this, even if [O~III]5007/H$\beta$ were as high as 2.5, then [O~III]/H$\alpha$ would be at most 0.85, and in P3, this would go as low as 0.38. Using the \texttt{dustmaps} code \citep{dustmaps}, we find little evidence of significant dust extinction, with values of E(B-V)$\sim$0.06 for all pointings in the nebula. The recent work by \cite{Ogle} presents deeper [O~III] and H$\alpha$ images, with a new photometric calibration that seems to resolve this discrepancy. Their estimations of the [O~III]/H$\beta$ ratio (derived directly from [O~III]/H$\alpha$) range from 0.8 to 2 in the areas with strong [O~III] emission. This is compatible with our spectroscopic measurements and is close to the spectroscopic result in \cite{Fesen23}.

We can also indirectly compare our JAST80 photometric data with the MEGARA line fluxes, using the same estimation of H$\alpha$/H$\beta$=2.86. We must also include [N~II]6583, which falls within the wavelength range of the $J0660$ filter. This results in an estimated flux in P1 $\geq$ (6.72 $\pm$ 0.88) $\times$ 10$^{-18}$ erg s$^{-1}$ cm$^{-2}$ arcsec$^{-2}$. The average surface brightness we measure in the $J0660$ detection is (5.50 $\pm$ 1.0) $\times$ 10$^{-18}$ erg s$^{-1}$ cm$^{-2}$ arcsec$^{-2}$, compatible with the spectroscopic value in P1 (assuming very low dust extinction, as mentioned before).

\begin{table*}[t] 
\caption{Receding velocities and fluxes of the different emission lines in the nebula, for all four pointings and both nights.}\label{tab:veloc_flux}
\centering 
        \begin{tabular}{c c c c c c c} 
        \hline\hline 
      Emission  & Pointing     &  \multicolumn{2}{c}{Velocity} & Velocity difference     & Flux & Width \\ 
         line       &         & \multicolumn{2}{c}{[km/s]} & [km/s]  & [10$^{-18}$ erg s$^{-1}$ cm$^{-2}$ arcsec$^{-2}$] & [km/s]\\ 
        \hline
 & & D1 & D2 & & D1 &\\
                \hline

        \multirow{4}{*}{H$\beta$}  &  P1     & $-27.9\pm$2.0   & $-29.1\pm$1.2 &  \multirow{4}{*}{$9.0\pm3.2$} & 1.72$\pm$0.28 & \multirow{4}{*}{$22.0\pm$1.4}\\ 
       &  P2     & $-31.3\pm$1.4    & $-34.1\pm$2.6  & & 1.64$\pm$0.16 & \\ 
        &  P3    &       & $-25.3\pm$2.7 & & 1.68$\pm$0.12 &\\ 
        &  P4    &        & $-30.2\pm$2.8 &  & 1.12$\pm$0.20  &\\
                   \hline
                          \multirow{4}{*}{[O~III]4959}  &  P1       &    & $-6.9\pm$4.7   &  &  &\\ 
       &  P2     & $-13.6\pm$2.7    &  &  & 0.56$\pm$0.12  & \\ 
        &  P3    &       &  &  &  &\\ 
        &  P4    &        &  &  & & \\
                   \hline
            \multirow{4}{*}{[O~III]5007}  &  P1   & $-7.7\pm$1.3    & $-12.4\pm$2.3  &  \multirow{4}{*}{5.4$\pm$1.9} &  2.44$\pm$0.24 &  \multirow{4}{*}{21.9$\pm$1.0}\\ 
       &  P2     & $-6.2\pm$1.1    & $-10.4\pm$1.9  &  & 2.04$\pm$0.16 &\\ 
        &  P3    & $-11.6\pm$1.4    & $-10.7\pm$2.8 &  &1.88$\pm$0.16 &\\ 
        &  P4    &        &  & & &\\
                   \hline

            \multirow{4}{*}{[N~II]6583}  &  P1   & $-20.1\pm$1.9    & $-19.7\pm$2.2   &  & 1.64$\pm$0.08& \multirow{4}{*}{27.8$\pm$1.5}\\ 
       &  P2     & $-29.4\pm$1.6    & $-27.1\pm$2.7   &   & 1.64$\pm$0.12 & \\ 
        &  P3    &       &   & &  &\\ 
        &  P4    &        & $-21.5\pm$3.1 &  &0.84$\pm$0.16 & \\
                   \hline

            \multirow{4}{*}{[S~II]6716}  &  P1   & $-20.2\pm$2.4    & $-20.4\pm$2.9   &  & 1.32$\pm$0.16  & \multirow{8}{*}{28.3$\pm$2.1}\\ 
       &  P2     & $-27.7\pm$2.5    &   & & 1.44$\pm$0.12 &\\ 
        &  P3    &       &    & &  &\\ 
        &  P4    & $-20.2\pm$2.4      &  &  & 1.12$\pm$0.12 &\\

            \multirow{4}{*}{[S~II]6731}  &  P1   & $-19.4\pm$1.9     & $-18.1\pm$3.7 
  &    & 0.96$\pm$0.12 &\\ 
       &  P2     & $-24.9\pm$1.6    &    &   & 1.00$\pm$0.08 & \\ 
        &  P3    &       &   &  &  &\\ 
        &  P4    & $-14.2\pm$3.9   & $-13.11\pm$5.3   &  & 0.88$\pm$0.16 &\\

                    \hline
        \hline 

        Sky line 5198~\AA & & & & & &  19.7$\pm$ 0.546     \\ 
         \hline 
       Sky line 6554~\AA & & & & & &  19.10$\pm$ 0.05     \\ 
        \hline 
        Sky line 6864~\AA & & & & & &  19.25$\pm$ 0.12     \\ 
                            \hline
        \hline 
\end{tabular}
\tablefoot{ Missing values are due to lack of detection (except for P3 in the three reddest lines, where no observation was taken). Fluxes are computed integrating the best-fitting Gaussian profile to each line. Note that line fluxes in the second night are considered less reliable due to flux calibration issues and thus not shown (see Sect. \ref{sec:results_emlineflux}). The velocities are presented in the barycentric reference system. The columns "Velocity difference" and "Width" show the median values of the nebula pointings (P1, P2, and P3, if available) for each specific emission line. The velocity difference reports the maximum range  and the median uncertainty, while the width value shows the median width and the typical dispersion of the values.}
\end{table*}

\subsubsection{Velocity derived from the emission lines}

\begin{figure}
\includegraphics[width=0.5\textwidth,keepaspectratio]{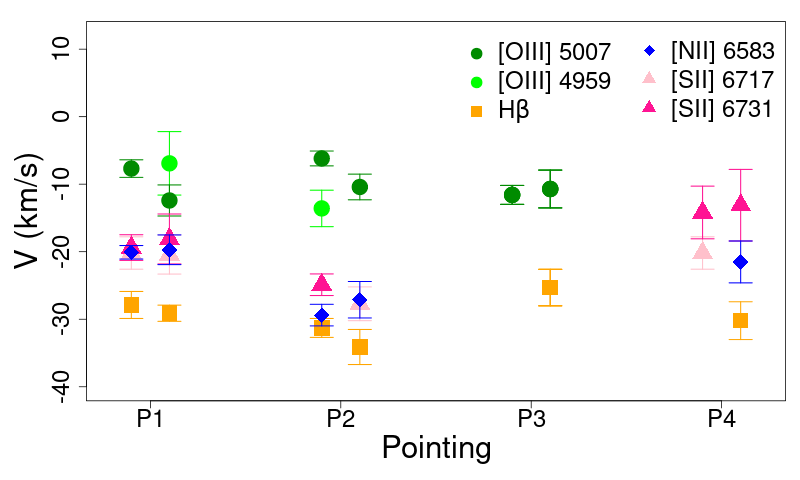}
      \caption{Receding barycentric velocities derived from different emission lines and pointings. For each pointing, the first column corresponds to results from the first night of observations, and the second one to the second night. Color code and lines missing as in Fig. \ref{fig:fluxes}. }
         \label{fig:veloc}
   \end{figure}

After performing the Gaussian fit to the emission lines, our code also outputs the difference between each measured line and its rest-frame wavelength, in units of km s$^{-1}$, for all lines and pointings on both days. The errors are very low: on the order of 2 - 3 km s$^{-1}$ for the nebula emission lines and below 1 km s$^{-1}$ for the much stronger sky emission lines. We used these sky lines as reference to perform the final correction on the nebular line velocities, considering that their velocity shift should be compatible with zero, since they originate in our atmosphere. Finally, we have then placed the nebular velocities in the barycentric system (see Appendix \ref{app:spec} for more details of the process). We report the velocities and their uncertainties for all the nebula emission lines that are successfully fitted by our code in Table \ref{tab:veloc_flux}, and show them in Fig. \ref{fig:veloc}.

 Finally, we can compare our measurements with previous spectroscopic observations of SDSO. Our pointing P1 is located just 90 arcsec west of the placement of Slit 2 in \cite{Fesen23}, allowing us to compare both measurements. Like them, we detect H$\beta$, [O~III]4959, and [O~III]5007. The heliocentric velocity we measured for [O~III]5007 on both nights ($-7.5\pm1.9$ km s$^{-1}$ the first one and $-6.7\pm$2.3 km s$^{-1}$ the second) are perfectly compatible with their value of $-$9.8$\pm$ 6.8 km s$^{-1}$, rendering more confidence to both our measurements. Their H$\beta$ determination ($+34\pm$14 km s$^{-1}$) is, in contrast, very different from ours ($-$31.4$\pm$1.2 km s$^{-1}$), a 4.7 $\sigma$ discrepancy considering their measurement. The lower S/N of their H$\beta$ line and its wider profile may explain part of the discrepancy, along with the fact that their H$\beta$ measurement only takes into consideration around half of the size along the slit used in their [O~III]5007 measurement (and while nearby, they do not overlap with our measurement).

However, our results on SDSO are not compatible with those in \cite{Amram23}. They measure a velocity of $-$96$\pm$4 km s$^{-1}$ fitting all lines they detect: a 16$\sigma$ discrepancy with our lowest velocity measured (H$\beta$). The source of this difference is not immediately obvious. The distances between our P1 region and their slit, 6.5 arcmin, is larger than that separating \cite{Fesen23} and P1, but it is still small compared to the distances between our pointings (almost 30 arcmin), and yet we find consistent velocities across the nebula. We must take into account that the resolution in \cite{Amram23} is between 2.4 to 2.9 times lower than ours, resulting in broader sky emission lines, contaminating the H$\alpha$, [N~II], and [S~II] emission lines (their Fig. 2). In addition, their [O~III] emission is contaminated by Sodium emission from urban street lights.

\subsubsection{Emission line widths}

In order to further investigate the physical properties of the gas in the nebula, we measured the width of the emission lines. More precisely, we report the standard deviation ($\sigma$) of the best-fitting Gaussian model for each line with the same code we used to measure the emission line velocities and fluxes. However, this measured width is the convolution of the intrinsic width of the line and the instrumental profile.

We study the width of the sky emission lines in our data to get a direct measurement of the instrumental broadening effect. Their intrinsic line width is very small, below $\sigma$=0.01 \AA~in all the sky lines that we detect \citep{Hanuschik03}, and therefore we can assume that the width we measure for them is essentially due to instrumental effects. We have measured the same sky emission lines mentioned in the previous subsection, and have obtained a very consistent width value, 19.30 $\pm$ 0.30 km s$^{-1}$. Separating by VPH, the difference is not striking: 19.70 $\pm$ 0.37 km s$^{-1}$ for the only sky line in LR-B, 5198 \AA, and 19.2 $\pm$ 0.16 km s$^{-1}$ for the three lines considered in LR-R. This is reasonably consistent with the theoretical values (20.95 km/s and 21.20 km/s) obtained from the nominal spectral resolution of MEGARA VPHs\footnote{\url{https://www.gtc.iac.es/instruments/megara/}}. In the following analysis we consider as instrumental width the values derived from the sky emission lines.

In Table~\ref{tab:veloc_flux} we also summarize the observed line widths, separating by different lines, but considering together for each line all the pointings in the nebula P1, P2, and P3 and both days. The uncertainties shown correspond to the dispersion in the width values measured for a particular line in the different pointings and in both days.

Finally, to de-convolve the instrumental effect from these widths, we apply the following equation:
$$\sigma_{int}=\sqrt{\sigma_{obs}^2-\sigma_{inst}^2}$$

where $\sigma_{int}$ is the intrinsic velocity dispersion, while $\sigma_{obs}$ is the observed one and $\sigma_{inst}$ is the instrumental dispersion. For $\sigma_{obs}$ values very similar (considering uncertainties) to $\sigma_{inst}$ we can only provide upper limits for $\sigma_{int}$. For H$\beta$ and [O~III], $\sigma_{int}$ must be less than 17 km s$^{-1}$ and 19 km s$^{-1}$, respectively, considering 3$\sigma$ limits. The [N~II] and [S~II] emission line widths are however significantly (more than 4$\sigma$) wider than the instrumental width, as traced by their nearby sky emission lines. Applying the equation, we obtain $\sigma_{int}$ =20.2 $\pm$ 2.1 km s$^{-1}$ for [N~II], and $\sigma_{int}$=20.7 $\pm$ 3.1 km s$^{-1}$ for [S~II].

\section{Discussion}
\label{sec:discussion}
In this section, we discuss and compare the properties we measure in SDSO with those of known astrophysical objects. We take into account both Galactic and extragalactic structures, and consider which ones are more and less likely to be the category that SDSO belongs to.

\subsection{Possible categories of astrophysical structures where SDSO may fall}

The main division that can be made between the different physical explanations for SDSO is between Galactic and extragalactic objects. If the nebula is located within the Milky Way, we consider that it could be a supernova remnant (SNR), a planetary nebula, a filament of diffuse ionized gas (DIG), or a bow shock due to stellar ejecta. These options were explored and discarded by \cite{Fesen23}, while recent work by \cite{Ogle} supports the latter option. More precisely, they propose that SDSO is caused by the bow shock as a result of the interaction between the local ISM and the ejecta from an old planetary nebula. We also consider the possibility that SDSO may be part of a filamentary structure in the diffuse ISM of the Milky Way.

If SDSO were an extragalactic object physically close to M31, \cite{Fesen23} proposed that it may be the result of a galaxy-wide ionized shock front due to the interaction of the circumgalactic medium (CGM) of M31 and the Milky Way. We also consider the possibility that SDSO may be an EELR associated with M31.

In Table \ref{tab:objects} we summarize the properties of SDSO alongside the typical properties of the different classes of objects that we consider as potential physical explanations for the nebula. In the following subsections we review the different sources of evidence about SDSO, and how these support or not the different possibilities about the physical nature of the nebula.

\begin{table*}[t] 

\caption{Properties measured in SDSO in this work, along the typical ranges that different categories or nebulae show in these properties.}\label{tab:objects}
\centering 

        \begin{tabular}{c c c c c c c c c} 
        \hline\hline 
      Properties  & SDSO     &  \multicolumn{5}{c}{Galactic nebula} & \multicolumn{2}{c}{Extragalactic nebula}   \\ 
      \addlinespace[2mm]
     & & H II & DIG & SNR & PN & Bow shock & EELR & CGM shock \\
\hline 
\hline           
\addlinespace[2mm]

     Radial Velocity & -6 to -30& \multicolumn{5}{c}{-300 to +300} &\multicolumn{2}{c}{-100 to -500} \\
      $[$km s$^{-1}$]  & & & & & & & & \\
      \toprule
      \toprule
      \addlinespace[2mm]
    Line width ($\sigma$) & $\leq$ 20 & $<$ 30 & $\leq$25 & $>$40 & 20 - 30& $>$25& 10 - 20 & >50 \\
    $[$km s$^{-1}$]  & ([O~III])& & & & & & &\\
      \midrule 
      \addlinespace[2mm]
    log$_{10}$([O~III]/H$\beta$)  & 0.04-0.15& -1 to 0.8 & -0.5 to 0.5  & -0.5 to 1 & 0.5 to 1.2 & -1 to 1 & $>$0.4 & -0.5 to 1\\
      \midrule 
      \addlinespace[2mm]
     log$_{10}$([O~III]/[O~II]) & $\sim$ -0.5 & -1 to 1 & -0.5 to 0.5 & -0.8 to 0 & 0.6 to 1.3 & - & 0.3 to 0.8 & - \\
      \midrule 
      \addlinespace[2mm]
    Electron density & $<$300 &  $<$600 &  0.02 - 20 & 200-10000 & $<$1000 & 0.1 - 1000 & 3-300 & - \\
     $[$cm$^{-3}$] & & & & & & & &\\

      \midrule 
      \addlinespace[2mm]
     Emission line & 6 arcmin &\multicolumn{5}{c}{5.2 arcsec to 10 deg} &\multicolumn{2}{c}{0.13 to 6.6 arcsec} \\
      separation & & & & & & & \\
    \hline 
\end{tabular}
  \tablefoot{In the first and last rows we only separate Galactic and extragalactic nebula, stating simply the radial velocity from the heliocentric system. In the last line, we assume a physical separation between regions emitting in different emission lines ranging from 0.5 to 25 pc, based on the literature (see Sect. \ref{sec:morpho}). The stated values show the apparent size when considering distances from 10 pc to 10 kpc (galactic nebula) and 770 kpc (M31 nebula). HII regions \cite{Nakajima}, DIG: \cite{Gonzalez-diaz24b,denbrokdig}, SNR: \cite{snrs,snrhowfesen, snrdens}, PNs: \cite{PNs,pndens}, Galactic shocks: \cite{green-ophiuchi,Ocker_shock,Schultz}, EELRs: \cite{Keel12}, Extragalactic shocks \cite{Rodriguez-Baras,weave,rich}}
\end{table*}

\subsection{Morphology}
\label{sec:morpho}

\begin{figure}
   \centering
   \includegraphics[width=0.45\textwidth,keepaspectratio]{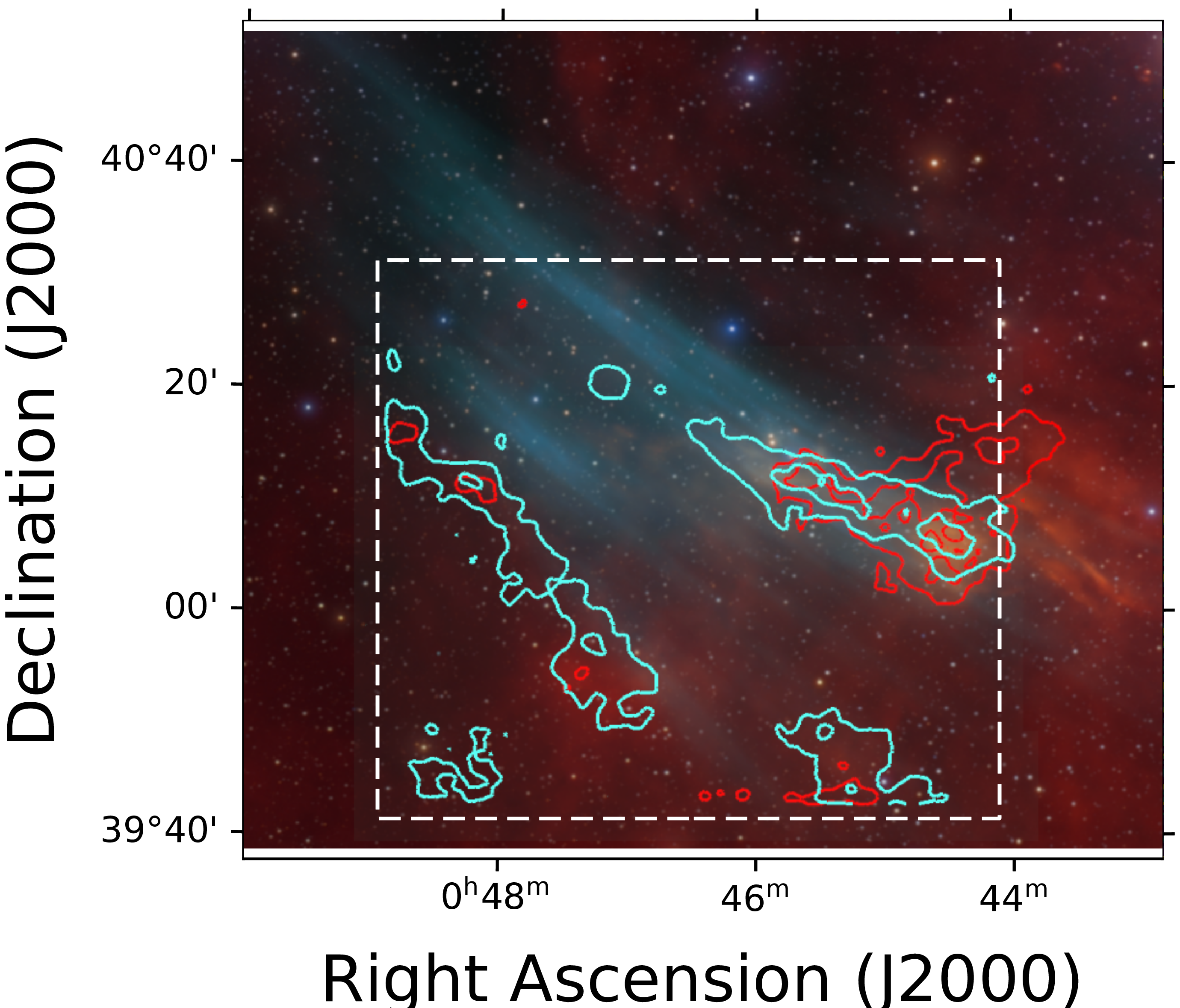}
      \caption{Contours of emission in the JAST80 observations on the nebula, for [O~II] (filter $J0378$) in cyan and for H$\alpha$ in red (filter $J0660$). The dashed-line square represents the region with full depth in the JAST80 images. In the background, the RGB composition shown in Fig \ref{fig:pointings} and presented in  \cite{Drechsler23}, with red tracing H$\alpha$ and cyan [O~III].}
         \label{fig:oii_oiii}
   \end{figure}

The main results identified in the OAJ narrowband imaging data are the near-parallel emission line stripes in the [O~II] 3727 emission image (traced with the $J0378$ filter). The position of these stripes does not match the position of the regions with emission in [O~III] as traced in \cite{Drechsler23}, which we show in Fig. \ref{fig:oii_oiii}.

The brightest one (on the right) partially overlaps with the brightest part of the main strand of the [O~III] nebula, and then, in its easternmost part, it separates slightly to the south. It makes a small angle of $\sim$ 17 deg with the [O~III] emission, and ends on the secondary structure of the main [O~III] strand. The fainter region in our [O~II] image is close to the fainter [O~III] strand, but also shows a significant separation toward the south. This separation reaches a minimum value of 6.5 arcmin between the brightest points of the regions in their nearest point. They both follow roughly the same south-east direction as the other [O~II] and [O~III] strands, also making a small angle (14 deg), but this time in the opposite direction.

To interpret the morphology of SDSO, we note that significant angular separation between the regions where the maximum flux of emission lines is located can be observed in nearby nebulae. This occurs only if they are close enough or large enough for its stratification structure to be resolved. In \cite{Barman}, they show an example with a separation of $\sim$ 4 pc. If we look at the distance between peaks of [O~III] and [S~II] emission, in \cite{pellegrini} these values can be higher, up to 20 to 25 pc. In one of the largest \textsc{H\,ii} regions in the local group (NGC604 in M33), with \textit{Hubble Space Telescope} (\textit{HST}) data (P.I. Barbá R.)\footnote{\url{https://archive.stsci.edu/proposal_search.php?id=10419&mission=hst}} presented in \cite{barba}, we measured a separation of $\sim$ 4 pc between [O~III] and [O~II] strands.

In the case of SDSO, even if the physical distance between the emission of the lines was in the large end of the range (25 pc), the resulting angular separation would be just 6.63 arcsec at the distance of M31, 500 times smaller than what we see in our images. Considering this range of physical separation (from 4 to 25 pc), in order to match the 6 arcmin we measure between the [O~III] and [O~II] emitting strands, the physical distance to the nebula would be between 2 and 14 kpc. These values would place the object well within the Milky Way. In consistency with these calculations, no EELR that we have seen referenced in the literature shows any measurable stratification in its ionization structure, since the expected separation would be on the order of milliarcseconds or less.

This separation can be seen in multiple Galactic structures, where the ionization structure can be resolved because of the shorter distance to them. An example is the SNR Veil Nebula in the Cygnus loop \citep{cygnus}, another is the Sh 2-174 HII region \citep{sharpless,Frew2008}, an ionized gas cloud with complex morphology that shares some kinematic and morphological properties with SDSO. This structure contains two near-parallel arcs emitting mostly in [\ion{O}{iii}], with a more amorphous H$\alpha$ emitting region, as seen in the image\footnote{https://www.astrobin.com/4mafcb/0/} by R. Shepard (red for H$\alpha$ and blue-turquoise for [O~III]). The angular separation of around 7 arcmin between those regions translates into 0.6 pc in physical separation, considering the distance to Sh 2-174 \citep{gaia}.

Finally, we can consider the general structure of SDSO in [O~III], as shown in the amateur images. It is very elongated and composed of several strands, with axis ratios around 10:1, depending on where the separations are placed. Its surface brightness distribution is smooth, without presenting areas much brighter than others, and without any clear limb brightening. However, large extragalactic ionized structures such as EELRs show a more intricate and rough morphology in [O~III], as can be seen in the HST images of \citet[][Fig.~8]{Keel12HST} and in \citet[][Fig.~1]{Keel15}. The physical scale covered by those images (from 14 to 60 kpc in size) is comparable to our Figure \ref{fig:pointings} at the distance of M31 (23 kpc). In the new data presented in \cite{Ogle}, there are additional [O~III] filaments (mentioned already in \citealt{Fesen23}) between the original SDSO and the disk of M31, around the star EG Andromedae. All these [O~III] filaments in SDSO are clearly different from the structure of the fountain-like object emitting in [O~III] located within the disk of M31, around RA 10.53 deg and DEC 40.9 deg. The latter shows a rougher, clumpier morphology than SDSO, more similar to what is seen in distant EELRs, further suggesting that the nebula is physically much closer to us than M31. In addition, the smoothness and elongation we see in the [O~III] images of SDSO can be seen in Milky Way structures, such as the previously mentioned Sh 2-174, planetary nebulae interacting with the local ISM like HFG1 (Fig. 4 in \citealt{Ogle} and \citealt{Chiotellis}), or other ISM filaments \citep{Hacar}. Regarding their size, ISM filaments span a wide range of physical sizes, from 0.1 to 200 pc \citep{Hacar}. If SDSO were a filament, given its $\sim$ 1.5 deg of angular size, this would place it between 3.8 pc and 7.6 kpc away. Although this is a very poorly constrained estimation, it shows that it could be a Galactic filament given its angular size. Compared to HFG1, its apparent size is roughly 6 times smaller than SDSO \citep{Chiotellis}, and therefore it would have to be 6 times farther away to be physically consistent. The resulting distance, $\sim$ 2.25 kpc, lies comfortably within the Milky Way.

Recent observations\footnote{\url{https://www.astrobin.com/ymtvkr/F/}} by amateur astronomers (The Deep Sky Collective\footnote{\url{https://deepskycollective.com/home}}) using multiple narrowband filters reveal "strands" in [S~II] emission that match the [O~III] structure well, supporting the hypothesis that different emission lines originate from the same physical structure despite velocity differences. This result has also been recently reported by \cite{Ogle}, after reprocessing the amateur astronomer data.

\subsection{Kinematics of the nebula}
\label{sec:discuss_pos}

\subsubsection{Receding velocity}
\label{sec:discuss_velocity}

The receding velocities of the emission lines in SDSO measured in the three MEGARA pointings ranged from $-8\, \rm{\kms}$ to $-30\, \rm{\kms}$ (Table~\ref{tab:veloc_flux}), with remarkable consistency for each ionic species across the three pointings P1, P2, and P3. These values are in agreement with observations of multiple ionized gas structures within our galaxy \citep[e.g., planetary nebulae in][]{floresduran}. Additionally, there are several examples of nearby nebulae in the Milky Way that also show velocity differences among emission lines from various ionic species belonging to the same structure, similar to those found in SDSO, for example between [\ion{O}{iii}] and [\ion{S}{ii}]. These include the Orion Nebula \citep{doi} and the interstellar medium (ISM) toward the Perseus arm (Fig.~9 in \citealt{Madsen06}), where the peak of emission in different emission lines differs by tens of $\kms$. Another comparison can be made with Sh 2-174 \citep{sharpless}. The heliocentric velocities of the emission lines in Sh 2-174 show some separation: $-11.2\, \rm{\kms}$ for H$\alpha$, $-7.3\, \rm{\kms}$ for [\ion{O}{iii}]5007, and $-15.0\, \rm{\kms}$ for [\ion{N}{ii}] \citep{Frew2008}.

On the other hand, the receding velocity of the stars and the gas in the region of M31 closer to the nebula (around half a degree away), ranges from $-400$ to $-250\, \rm{\kms}$, based on optical spectroscopy or HI observations \citep{Opitsch18,Corbelli10}. This represents a significant range of $>220\, \rm{\kms}$, in contrast with the measured velocities in SDSO, which would show nearly constant projected velocities over a region of $\sim 6\, \rm{kpc}$ in size if located at M31's distance ($770\, \rm{kpc}$).

Finally, the discrepancy between the velocity of SDSO and M31 is very rare in the observed EELRs, where they show, in almost all cases, consistent velocity with the rotation curve of their host galaxy (Fig.~12 in \citealt{Keel12}). In addition, some EELRs show significant velocity gradients over smaller distances than SDSO (e.g., Mkn 883, Teacup, or UGC 1185 in the Lick data in Fig. 12 from \citealt{Keel12}) while some others do not (SDSS 2201+11, Mkn 266, or NGC 5972 in Fig. 12 from \citealt{Keel12}).

\subsubsection{Velocity dispersion}
\label{sec:res_vel_dis}
The widths of the emission lines measured in the nebula are typically narrow, with velocity dispersions of $\lesssim 30\, \rm{\kms}$ for [\ion{O}{iii}]5007, H$\beta$, [\ion{N}{ii}], and [\ion{S}{ii}] (Table~\ref{tab:veloc_flux}). With respect to the velocity properties, the nebula shows consistent velocity dispersions for all the lines across the three MEGARA pointings. These low values are compatible with photoionized nebulae such as \textsc{H\,ii} regions \citep{weave}. For example, the emission line widths in the Sh 2-174 nebula are $\sigma \sim 11\, \rm{\kms}$, consistent with our measurements for [\ion{O}{iii}]5007 and H$\beta$. On the other hand, the narrow width of the lines is inconsistent with most shock-ionized nebula, according to the available literature. For example, ionized gas in all SNRs analyzed so far shows $\sigma >40\, \rm{\kms}$ \citep{points,Vicens-Mouret,DuartePuertas}. Galaxy-wide shocks, which are often found in galaxy interactions and clusters, present velocity dispersions much higher than 20 km s$^{-1}$ \citep{rich, weave}.

The small width of the emission lines also challenges possible explanations involving planetary nebulae or Wolf-Rayet stars. These objects present expanding gas shells, with typical double peaked line profiles and expansion velocities of a few tens of $\kms$ (e.g., \citealt{Chu91,Montoro-Molina23}).

Using the emission line widths, we can provide a rough estimate of the total mass and density of the nebula under the assumptions of uniformity and virial equilibrium. To estimate $k$, the scaling factor in the virial theorem, we assume an isothermal (smooth and diffuse) system, which results in $k=1$. The area of the nebula is approximated as 702 arcmin$^2$, modeled as a trapezoid with bases of 60 arcmin and 18 arcmin and a height of 18 arcmin. Assuming that the depth of the nebula is equal to its height, we compute a total volume of $1.42 \times 10^{11} \, \mathrm{pc}^3$ if the nebula is located at a distance of 770 kpc. For a velocity dispersion of $10\, \rm{\kms}$ (an approximation to that of [\ion{O}{iii}] and H$\beta$), the resulting gas mass is $7.5 \times 10^8\, \rm{M_\odot}$. The corresponding gas density would be $n_\text{HII} \sim 0.02\, \mathrm{cm^{-3}}$, comparable with low-density gas in the Milky Way ISM ($0.03$–$0.08 \, \mathrm{cm}^{-3}$; \citealt{Haffner}), while other extragalactic ionized structures such as EELRs typically have larger densities of $10$--$100\, \mathrm{cm}^{-3}$ \citep{Keel12}. Alternatively, adopting similar parameters for a distance of 7 kpc within the Milky Way halo, the total volume would decrease to $1.1\times 10^5 \, \mathrm{pc}^3$, and the gas mass would drop to $6.8 \times 10^5 \, M_\odot$. The corresponding density would be $n_\text{HII} \sim 260\, \mathrm{cm}^{-3}$, in agreement with giant filaments in the ISM \citep{Hacar}, and also with the electron density estimated from the [\ion{S}{ii}] doublet (section~\ref{sec:discus_emilineflux}). This derivation presumes that the object is dynamically relaxed. Even if the assumption is not true, the mass estimations we obtain are still valuable as an order-of-magnitude approximation.

Overall, the kinematic properties of SDSO favor a Galactic origin for the nebula. Its measured velocities and line widths are inconsistent with the expected values for a structure associated with M31, while being entirely consistent with known Galactic photoionized gas structures.

\subsection{Emission-line ratios}
\label{sec:discus_emilineflux}

The [\ion{O}{iii}]5007/H$\beta$ ratios measured in the three pointings on the nebula (P1, P2, and P3) range from 1.1 to 1.4. These values are low for planetary nebulae \citep{PNs}, but consistent with several Galactic structures, such as \textsc{H\,ii} regions \citep{Frew10}, SNRs \citep{snrs} and the DIG \citep{Gonzalez-diaz24b}). In contrast, the ratios measured in EELRs in distant galaxies are significantly higher due to AGN ionization, almost always above 2.5 \citep{Keel12,Keel15,Keeletal24}. Galaxy-wide shocks present varied [O~III]/H$\beta$ ratios, some cases higher than our measurements \citep{Rodriguez-Baras}, others similar or lower \citep{weave}.

Without accurate H$\alpha$ flux measurements, we derived the [\ion{N}{ii}]/H$\alpha$ and [\ion{S}{ii}]/H$\alpha$ ratios using the H$\alpha$/H$\beta$ ratio of 2.86 that was discussed in Sect. \ref{sec:results_emlineflux}. 
Considering this, we obtain mean values of $\log_{10}$([\ion{N}{ii}]/H$\alpha$)= $-$0.313$\pm$0.038 and $\log_{10}$([\ion{S}{ii}]/H$\alpha$)= $-$0.469$\pm$0.038 for P1 and P2 pointings. These ratios place SDSO among most of the \textsc{H\,ii} regions and Wolf-Rayet nebulae in \citet{Frew10}, and are marginally consistent with those of planetary nebulae or SNR (Figures 5.2 and 5.3 in \citealt{Frew2008}). They are consistent as well with DIG values measured in external galaxies \citep{Gonzalez-diaz24b} or EELRs.

Finally, we can consider the [O~II] we measured in the JAST80 photometry (see Sect. \ref{sec:result_photom}), resulting in a [O~III]/[O~II] ratio of around 0.3. This value can be reached in typical galaxies \citep{Nakajima}, and also in DIG \citep{Galarza,Zhang}, but it is very different from the values measured in EELRs, typically from 2 to 6 \citep{Lintott09,Keel12}.

\subsection{Physical nature of the nebula}

In the previous sections, we have discussed the available physical measurements of the [O~III] nebula, and compared them with the properties of other classes of objects to constrain its location and characteristics. In this section, we review and integrate all the previous discussions.

The kinematical evidence strongly supports the Galactic nature of SDSO. First, the receding velocity we measure is not compatible with the velocity of gas and stars in M31, especially in the region closer to the nebula. In addition, the velocities measured for a given emission line across all pointings are remarkably consistent. This is not surprising if the nebula were within our Galaxy, and the distances between pointings were on the order of tens of parsecs, but it would be rare for an extragalactic cloud of gas across several kiloparsecs.

The emission lines are narrow, with velocity dispersion below 20 km s$^{-1}$ in H$\beta$ and [O~III]. This makes it very unlikely that the nebula is part of a SNR or stellar ejecta. Shock ionization, which has been considered the most likely source of ionization for SDSO in \cite{Fesen23} and \cite{Ogle}, is very likely incompatible with this result, since it requires velocities greater than 80 km s$^{-1}$ (\citealt{Kewley} and references therein). It is very hard to find previous evidence of gas ionized by shocks showing velocity dispersion below 20 km s$^{-1}$ especially galaxy-wide, which are often found in galaxy interactions and clusters. Nevertheless, a possible scenario that can reproduce the line ratios and the width of the lines would be a combination of low velocity shocks and photoionization \citep{Ali}.

The width of the emission lines also allows us to make rough estimates of the total mass and density of the nebula. If it were located at the distance of M31, it would be very massive (almost $\sim$ 10$^8$ M$_{\odot}$), requiring a very strong ionizing source like an AGN, which is absent in M31 (e.g., \citealt{Peng}). Moreover, it would have a very low density ($\sim$ 0.03 cm$^{-3}$), in contrast with the electron density of EELRs, which is 10$^2$ - 10 $^4$ times higher \citep{Keel12}. In contrast, if it were within the Milky Way, its derived properties (mass $\sim$ 10$^{5}$ M$_{\odot}$ and density $\sim$ 10$^{2}$ cm$^{-3}$) would be more in line with other known objects, like giant ISM filaments \citep{Hacar}. 

To provide additional support to the hypothesis that SDSO is an extragalactic ionization shock front, \cite{Fesen23} present evidence of high velocity clouds (HVCs) of neutral HI gas on the same area of the sky as the nebula. Nevertheless, the velocity of the clouds (-175 km s$^{-1}$) is very inconsistent with that of SDSO, and \cite{Fesen23} do not provide a potential explanation for the discrepancy. In addition, there are other HVC around M31, at similar velocities, and no [O~III] emission is associated with them, which hints that the spatial correlation of an HVC and SDSO may just be a coincidence. Although some studies do not reach low velocities, \cite{Kerp} reaches -25 km s$^{-1}$, above some of our velocity measurements in SDSO for H$\beta$, [S~II], and [N~II]. According to them, there is no evidence of HI cloud associated to M31 in the area where the SDSO is located (their Figure 7).

The measured emission line ratios are compatible with multiple Milky Way structures (ISM filaments, planetary nebula, \textsc{H\,ii} regions), with the [S~II]6716/[S~II]6731 ratios indicating an electron density below 300 cm$^{-3}$ \citep{McCallsii}. In contrast, the [O~III]5007/H$\beta$ ratio is very low compared with the vast majority of EELRs, which are one of the few known large ionized objects around external galaxies. In addition, these structures may show characteristic features in other parts of the electromagnetic spectrum. The original Hanny's Voorwerp has been identified in radio data from the LOw Frequency ARray (LOFAR, \citealt{lofar_hanny}). Visual inspection of the LOFAR data in SDSO \citep{Shimwell22} shows no emission, an additional piece of evidence against its EELR-like nature. Other negative evidence includes the lack of significant AGN activity in M31 (found in all known EELRs), the lack of significant galactic interaction (found in all EELRs in \citealt{Keeletal24}), and the smoother morphology of SDSO compared to EELRs (as discussed in Sect. \ref{sec:morpho} and seen in \citealt{Keel12HST}). Taking into account all the available evidence, we consider it extremely unlikely that SDSO is an EELR.

Finally, the angular separation between the emission in [O~II] and [O~III] can be observed in other gas clouds, but it would be too small to be visible if the nebula were located at extragalactic distances, and there are in fact no reports of it in EELRs or galaxy-wide shocks. The separation between ionized species can also be seen in the Sh 2-174 Galactic nebula discussed earlier, with [O~III] filaments separated from H$\alpha$ diffuse emission.

Considering the evidence we have gathered and presented, our most compelling explanation is that the nebula belongs to the ISM of the Milky Way, probably to its diffuse phase. Given its morphology, the nebula may be considered an ISM filament. As stated in \cite{Fesen23}, most of these filaments are detected in radio or infrared, with only a few having optical emission line observations, almost always with H$\alpha$ \citep{McCullough,West}. Nevertheless, this may not be a physical necessity, but just due to a lack of spectroscopic follow-up observations on Galactic filaments targeting other emission lines. Deep photometric narrowband surveys other than in H$\alpha$ are certainly missing, as proven by \cite{Drechsler23}, discovering such a large structure in an area of the sky surveyed for more than 100 years. Despite that, there are some indications that these H$\alpha$ filaments may have some [O~III] emission, such as some of the high Galactic latitude structures in \cite{Madsen06}. 

Moreover, the lack of change in velocity across SDSO is also consistent with the velocity gradients measured in Galactic filaments. In \cite{Hacar}, typical values of gradients and lengths can result in velocity shifts of at most a few km s$^{-1}$, and as low as less than 1 km s$^{-1}$, as also shown in \cite{Wang}.

Finally, some evolved star ejecta may also create filament-like features, with separated [O~III] and H$\alpha$ emitting regions, like those shown in \cite{Garnett,Gruendl,Stupar}. However, cases with [O~III] detached from H$\alpha$ are often caused by the expansion of a shock within a lower density medium, an effect that can be seen as well in planetary nebulae \citep{guerrero}. Nevertheless, in those cases, the morphology is often very sharp, in contrast with SDSO.

In the recent paper by \cite{Ogle}, the authors propose that SDSO is in fact a shock-ionized gas cloud that forms a "Ghost Planetary Nebula" within our Milky Way. The object proposed as the central star of the planetary nebula is EG\,Andromedae, a symbiotic star moving hypersonically with respect to its surrounding ISM, with a proper motion pointing toward SDSO. In their model, the gas expelled by the star is ionized by shocks because of the velocity of its interaction with the ISM. This is supported by the measured line ratios, the morphology of the [O~III] and H$\alpha$ clouds, and the position and proper motion of EG Andromedae. Their model opens a new potential scenario to understand the properties of SDSO in a Galactic context, explaining the morphological configuration and the observed line ratios. However, some implications of this model are hard to reconcile with our measurements. The radial velocity of the proposed star that would have originated the planetary nebula, EG Andromedae, is -95 km s$^{-1}$. According to \cite{Ogle} this, coupled with a residual expansion velocity of 20  km s$^{-1}$, results in an expected velocity of around 115 km s$^{-1}$ at the head of the bow shock. This region should be the one with the strongest [O~III] emission in the nebula. In contrast, the measured velocities in all pointings (including the region with the strongest [O~III] emission, P1) are markedly different, being less negative than -15 km s$^{-1}$ in all [O~III] lines. The expected velocity shift toward more positive values when moving away from the "head" in \citet{Ogle}, reaching -13 km s$^{-1}$ in the sonic line of the bow shock, is not observed. We found no clear gradient in velocity between the three pointings observed, with remarkable consistency in [O~III] velocity across the nebula.

In the \cite{Ogle} model, the gas ionization is due to a shock moving at $\sim$ 100 km s$^{-1}$, with no contribution from photoionization. A shock with this velocity seems hard to reconcile with the velocity dispersions observed in the emission lines, which are low ($<$20 km s$^{-1}$ in the [O~III] lines). Bow shock velocity dispersions may be close to $\sim$ 25 km s$^{-1}$ for a 100 km s$^{-1}$ shock only in very specific geometric configurations \citep{Schultz}. Thus, accommodating the observed kinematics within the proposed framework would likely require an unusually complex geometry.

In addition, other characteristics of the model by \cite{Ogle} may lead to a contradictory scenario. The high velocity of the star with respect to the ISM would imply an intense loss of mass from the AGB wind shell as it gets dragged away. This would reduce the amount of mass and mechanical energy available for the outflow, and make it very complex for it to reach the current size and separation. Due to this, the nebula would stall against the ISM, and the star would be located much closer to SDSO, or even overtake it. Finally, while the \cite{Ogle} model is promising and offers a possibility to explain SDSO properties within the Milky Way context, we find that some further refinement would be required to reproduce the observed gas kinematics and other considerations discussed above.

The nebula shows mixed properties, some at the boundary between different classes of objects, and no clear ionizing source has been identified. It is important to note that, even if the object is ionized by a compact source, it may be very complex to clearly identify it if the object is very close to the solar system. The detailed discussion in \cite{Fesen23} shows that a strong ionizing source is unlikely to be located less than 10 degrees away from SDSO. However, if the object was very close to Earth (less than 60 parsecs) even if it was close to the nebula (10 parsecs) the projected distance would be well above 10 degrees. In addition, since our estimation of the [O~III]5007/H$\beta$ ratio is lower than \cite{Fesen23}, a colder object than what they considered could be enough to ionize SDSO. In any case, the absence of a clearly determined source for the ionization is common in gas filaments in the Milky Way and in the DIG in external galaxies. For the latter, multiple sources have been proposed: leaking photons from OB star associations; low mass, hot, evolved stars (HOLMES, \citealt{perez-montero}); shocks; cosmic rays, etc. \citep{Gonzalez-diaz24b}. Considering other studies of the DIG, such as \cite{Weber}, the line ratios we measure are consistent with that explanation ([O~III]/[O~II], [O~III]5007/H$\beta$, [S~II]/[N~II]). Other studies, such as \cite{Zhang,Koutsoumpou}, consider cosmic rays as an additional source for DIG ionization, and show line ratios compatible with those we measure in the nebula. Some DIG ionization methods discussed here may be the additional ionizing source responsible for the [O~III] emission we found in SDSO, a relatively uncommon feature in a typical Milky Way filament. 

Overall, we consider that the most likely scenario is that SDSO is a Galactic nebula, at least partially photoionized. With the data gathered up to now, we cannot determine precisely to which class of objects the nebula belongs to, but it is very likely to be located within the Milky Way and not physically close to M31. 

\section{Conclusions}
\label{sec:conclusions}
In this work, we have presented new observations, both spectroscopic and photometric, on the recently discovered large [O~III] nebula along a line of sight close to M31. The spectra consisted of four pointings of the MEGARA IFU, mounted at the 10.4-m GTC, while the photometry was obtained with narrowband filters tracing [O~II] and H$\alpha$ and [N~II] emission lines with the JAST80 telescope at the OAJ. The analysis of the data provides clear evidence that the object is not physically close to M31, but it is instead within our Milky Way. The key pieces of evidence are:

\begin{itemize}

    \item The radial velocity of all emission lines measured in all MEGARA pointings within the nebula show values less negative than $-$40 km s$^{-1}$, very different from the receding velocity of M31, $-$300 km s$^{-1}$

    \item The receding velocities measured for a given emission line in the different pointings on the nebula are very consistent, within a few km s$^{-1}$. This is normal in Milky Way structures of a few tens of parsecs of size, but uncommon in extragalactic gas extending over several kiloparsecs.

    \item The velocity dispersion of the emission lines (below 20 km s$^{-1}$ for H$\beta$ and [O~III]5007) is compatible with photoionized nebulae but too low for shock ionization. Considering this line width and the virial theorem, if SDSO were located in the Milky Way, the mass and density of the nebula would be consistent with giant ISM filaments. However, if it were at the distance of M31, it would be very massive and with extremely low density, inconsistent with EELRs, the most well studied extragalactic [O~III]-emitting gas clouds.

    \item The [O~III]5007/H$\beta$ line ratios we measure in the nebula pointings (from 1.12 to 1.4) are consistent with several Galactic structures, but they are much lower than what is typical in EELRs. The lack of AGN and galaxy interaction in M31 are also evidence against the EELR nature of SDSO.

    \item In contrast with previous analysis, we identify clear hydrogen emission (H$\beta$) across the nebula. This implies that the  strong [O~III]/H$\alpha$ (and [O~III]/H$\beta$) ratios in the nebula are lower than previously considered. This makes the nebula more compatible with several structures within our galaxy that may have been ruled out otherwise, such as filaments or DIG.

    \item In the narrowband images tracing [O~II] we detect two strands of emission near the [O~III] nebula, with similar orientation, but with a $\sim$ 7 arcmin shift. This is easy to explain if the [O~II] strands are part of the nebula, and thus the separation is just a few parsecs, making SDSO a Milky Way object.
    
\end{itemize}

We have presented compelling evidence demonstrating the location of the nebula within the Milky Way, but its precise nature and the class of objects it belongs to is less clear. Considering the line ratios ([O~III]5007/H$\beta$, [S~II]/[N~II], [S~II]6716/[S~II]6731), the lack of an evident source for the ionization, its filamentary morphology, and the width of the emission lines ($<$ 20 km s$^{-1}$ for [O~III] and H$\beta$), the most fitting category to place the nebula is that of Galactic ISM filaments.

Further spectroscopic analysis, on other locations in and around the nebula and covering bluer wavelengths, will help shed light onto its physical nature. The discovery and analysis of this object highlight the importance of observing wide fields of the sky with narrowband filters, as shown in \cite{Fesen24}. Large-scale surveys of the sky in narrowband filters, such as J-PLUS \citep{Cenarro19C} and especially the deeper J-PAS \citep{Bonoli} will open a new window into the emission-line sky.

\begin{acknowledgements}

ALC would like to thank Asier Castrillo and Raúl González-Díaz for useful discussions. The authors would like to thank the team of OAJ operators responsible for the photometric observations, as well as the GTC team responsible for the spectroscopic observations. The authors acknowledge funding by the Governments of Spain and Aragón through their general budgets and the Fondo de Inversiones de Teruel. ALC acknowledges funding support by the European Union - NextGenerationEU through the Recovery and Resilience Facility program Planes Complementarios con las CCAA de Astrofísica y Física de Altas Energías - LA4. The authors acknowledge funding by the Governments of Spain and Arag\'on through their general budgets and
the Fondo de Inversiones de Teruel. ALC acknowledges funding support by the European Union - NextGenerationEU through the Recovery and
Resilience Facility program Planes Complementarios con las CCAA de
Astrof\'{\i}sica y F\'{\i}sica de Altas Energ\'{\i}as - LA4. 
ALC, JAFO, AHC
acknowledge financial support by the Spanish Ministry of Science and Innovation (MCIN/AEI/10.13039/501100011033), by ``ERDF A way of making Europe'' and by ``European Union NextGenerationEU/PRTR'' through the grants PID2021-124918NB-C44 and CNS2023-145339; MCIN and the European Union -- NextGenerationEU through the Recovery and Resilience Facility project ICTS-MRR-2021-03-CEFCA. RIS acknowledges financial support from the Spanish Ministry of Science and Innovation through the project PID2022-138896NA-C54. HVA is supported by the grant PTA2021-020561-I, funded by MICIU/AEI/10.13039/501100011033 and by ESF+.
AE acknowledges the financial support from the Spanish Ministry of Science and Innovation and the European Union - NextGenerationEU through the Recovery and Resilience Facility project ICTS-MRR-2021-03-CEFCA.
This work was partly done using GNU Astronomy Utilities (Gnuastro, ascl.net/1801.009) version 0.23. Work on Gnuastro has been funded by the Japanese Ministry of Education, Culture, Sports, Science, and Technology (MEXT) scholarship and its Grant-in-Aid for Scientific Research (21244012, 24253003), the European Research Council (ERC) advanced grant 339659-MUSICOS, the Spanish Ministry of Economy and Competitiveness (MINECO, grant number AYA2016-76219-P) and the NextGenerationEU grant through the Recovery and Resilience Facility project ICTS-MRR-2021-03-CEFCA and the Spanish research agency grant PID2021-124918NA-C43.

This work made use of Astropy\footnote{\url{http://www.astropy.org}}: a community-developed core Python package and an ecosystem of tools and resources for astronomy \citep{Astropy22}.
\end{acknowledgements}

\bibliographystyle{aa}
\bibliography{bibm31}

\begin{appendix} 
\section{Extended description of the photometric observations and data analysis}
\subsection{Photometric observations and data reduction}
\label{app:data}

To facilitate the data reduction process and the detection of LSB structures, we used a large dithering pattern of about half a degree when planning the photometric observations of SDSO with JAST80. We implemented this by defining four separate pointings, each one with an additional, smaller dithering pattern of 10 arcsec. Since the maximum exposure time in T80cam is 600 s (10 minutes), we requested 6 individual observations in each pointing for the $J0378$ image, and 3 for $J0660$. However, given the short time frame available to secure all observations due to M31 visibility, one of the pointings in the $J0660$ was not observed, and two exposures in one of the $J0378$ pointings are missing as well. We therefore obtained 22 individual 600 s images in $J0378$ for a total of 13200 s (3.67 h) of exposure in the central region, and 9 images for a total of 5400 s (1.5 h) of exposure in $J0660$. The different pointings performed only overlap in the central area, of roughly 56 arcmin $\times$ 50.3 arcmin (0.78 deg$^2$), which covers most of the [O~III] nebula, as shown in Fig. \ref{fig:pointings}.

The raw exposures were processed by the data processing and archival department (DPAD) at CEFCA. The standard J-PLUS pipeline was used to perform the bias and flat field corrections and flux calibration. In addition to that, during the sky background correction step, special care was taken in order to avoid the removal of potential LSB structures on the images.
To do this, the Gnuastro \texttt{NoiseChisel} program \citep{akhlaghi2015,akhlaghi2019} was used to accurately detect all significant objects in the image without any parametric fitting to their profiles, and to estimate a nonparametric model of the sky background. With this software it is possible to detect very faint signal from LSB features (i.e., the faint nebulae) and mask them, to keep them in the final images and avoid incorrectly subtracting them as the sky background.
The final models of the sky background emission were estimated by fitting the \texttt{NoiseChisel} sky tiles (before interpolation) with 2-dimensional Chebyshev polynomials (order=3 for $J0378$, and order=4 for $J0660$) to the previous sky background models.
After subtracting the sky background emission, all individual frames were resampled into the same common grid and sigma-clipped and mean-stacked to obtain the co-added images.

In order to remove the contamination of the scattered light from the very bright stars on the co-added images, the extended point spread function (PSF) was estimated. The method we followed is described in detail in \cite{infantesainz2020}. In short, very bright stars (\citealt{gaia}, $G<$ 10\,mag) were processed and stacked to measure the far extended wings of the PSF (up to a radius of $\sim$ 3\,arcmin). Fainter and non-saturated stars ($G =$ 10 -- 14\,mag) were used for the core of the PSF, and both of these sources were combined into a single extended PSF model. This model was then subtracted from the bright stars ($G<$ 14\ mag) in the images to remove their scattered light and avoid contamination of LSB structures.
For fainter stars ($G>$ 14\ mag) this procedure was not applied, since
their brightness does not extend very far away. They were simply detected and masked instead. Finally, with the goal of increasing the S/N ratio of faint structures, the pixel scale of the images was increased to 20\ arcsec pix$^{-1}$.

\subsection{Photometric data analysis}
\label{app:photom_anal}
In these observations we targeted very faint structures, and we pushed the data to reach very low surface brightness. This resulted in the reveal of some systematic effects, and therefore the estimation of the significance of the detections requires special care. A simple analysis, pixel by pixel, of the S/N would overestimate it, overlooking the large-scale systematic effects (such as the circular dark regions visible across the image in Fig. \ref{fig:oaj_photom_data}, corresponding to the removal of stars). To overcome that, we performed flux measurements in 10000 apertures with the same shape and size of the ones we have used to make the measurements, but randomly placed all around the image, using the \texttt{--upperlimit-sigma} column of Gnuastro's \texttt{MakeCatalog} program \citep{makecatalog}. This process then takes the width of the distribution of those 10000 fluxes as an upper limit to the uncertainty in the flux values in our regions of interest. Depending on where we allow these random apertures to be placed, the S/N gives different values. If we only avoid placing them on the aperture of the region we are considering (thus assuming that everything but the region of interest is noise), we obtain values of S/N = 5.1 in the first region and 5.9 in the second. If we mask every detection in the image using \texttt{NoiseChisel}, this S/N value goes up to 13.3 and 13.4, respectively. The fainter structure shows higher S/N due to its orientation, which is not aligned with the linear structures that seem to cross the whole image. In any case, this test demonstrates that these structures are confidently detected above the 5$\sigma$ level. In the H$\alpha$ image we detect a roughly triangular region of $\sim$ 24-27 arcmin in size, and with the same method we estimate the S/N to range from 5.3 to 6.7, confirming the reality of the structure.
\section{Extended description of the spectroscopic data analysis}
\label{app:spec}
\subsection{Data processing}
The spectroscopic data were reduced using the MEGARA data reduction pipeline, \texttt{megaradrp}, version 0.12.0 \citep{megara}. The procedure starts by subtracting the bias level from the raw FITS files, and then the individual spectra from each fiber have to be traced across the 2D images. This is done using the flat observations, where the trace of each fiber is very clear. Then, the flux from each fiber is calibrated in wavelength, using arc-lamp observations. Finally, each fiber is corrected from the differential sensitivity across wavelength with a flat-field correction. For the final results, we combined the three exposures that were taken for each nebula pointing with each VPH. The only exceptions were the LR-B observations of the P1 pointing. Those were performed early after twilight in both nights, and this led to the contamination of the spectra by sunlight, showing significant absorption lines that prevented us from accurately measuring the emission lines of the nebula, especially [O~III]4959 and H$\beta$. Therefore for this pointing, in both nights, we only used the third frame taken, which showed almost no sunlight effect.

The only difference in our data reduction process from the standard is that we did not perform sky subtraction. In MEGARA this process is performed by removing from the science data the flux measured by dedicated fibers located at $\sim$ 1.75 - 3 arcmin from the IFU. Given the large size of our target, those sky fibers are contaminated by the nebular emission, therefore using them to subtract the sky would remove the very emission lines we intend to detect. Our original plan to perform sky subtraction was to use the pointing outside the footprint of the nebula (P4), but those data were also contaminated by nebular emission lines, and therefore we performed no sky subtraction. Since we are expecting to only detect the emission lines from the nebula and not the continuum, our only concern are sky emission lines. They do not affect the nebular lines in the blue range of the spectrum, and only contaminate the H$\alpha$ in the red range (we address this in Sect. \ref{sec:results_emlineflux}). We also explored the possibility of detecting variations of emission within each pointing, by dividing the field into 9 equal squares arranged in a 3x3 pattern, but found no significant changes.

\subsection{Spectroscopic data analysis}

The spectroscopic observations were repeated in two separate nights, but for the second one there was no flatfield calibration available. Using the flat from the night after that one results in a poor calibration, presenting large-scale fluctuations in the flux values across wavelength, with significant slopes in regions that showed none on the first night. This, as we have shown in Sect. \ref{sec:discuss_velocity}, does not affect the wavelength calibrations, since we obtained remarkably similar velocity values for the same lines in both days. Even considering this issue, the flux values are reasonably consistent from the first night to the second night in most cases (13 out of the 16 lines detected in both nights are compatible within the uncertainties).

For the latter, the velocity shift should be compatible with zero in all cases, since sky lines come from our own atmosphere. Nevertheless, we measure differences much larger than the errors, from -5.9 km s$^{-1}$ to +3.5 km s$^{-1}$. These shifts are not random but correlated, being for example all positive on the first night and negative on the second night, and similar for different lines in each specific pointing. This indicates that the wavelength calibration is not stable enough across pointings or nights to directly compare the velocity shifts of the nebula emission lines. In order to correct for this effect, we have simply subtracted from all the velocities of the nebula emission lines, the sky emission line shifts in that specific pointing and night. For the blue grism, we have used the 5198 \AA~sky emission line since it is the only strong one detected. For the red grism, we have used the median shift of three strong sky lines that cover the wavelength range we are interested in: those at 6533.05~\AA, 6553.63~\AA, and 6863.97~\AA.

Finally, we have corrected the velocities to place them in the heliocentric or barycentric system. The corrections are near-identical for the two nights, $-$25.89 km s$^{-1}$ the first one and $-$25.91 km s$^{-1}$ the second one, using the routine \texttt{radial\_velocity\_correction} from \texttt{astropy} \citep{Astropy13,Astropy18,Astropy22}.

\end{appendix}
\end{document}